# Helicity-selective Raman scattering from in-plane anisotropic α-$MoO_3$

Shahzad Akhtar Ali[1], Abdullah Irfan[1], Aishani Mazumder[2], Sivacarendran Balendhran[3], Taimur Ahmed[2,4,5], Sumeet Walia[2] and Ata Ulhaq[1]*

[1]Physics department, Syed Babar Ali School of Science and Engineering,
Lahore University of Management Sciences LUMS, Sector U, DHA, 54792 Lahore, Pakistan

[2] School of Engineering, RMIT University Melbourne, Melbourne, Australia.

[3]School of Physics, The University of Melbourne, Melbourne, Australia.

[4]Functional Materials and Microsystems Research Group and the Micro Nano Research Facility
RMIT University Melbourne, Melbourne, Australia.

[5]Pak-Austria Fachhochschule: Institute of Applied Sciences and Technology, Haripur, 22620, Haripur, Pakistan.

*Corresponding email: ata.haq@lums.edu.pk

## Abstract

Hyperbolic crystals like α-$MoO_3$ can support large wavevectors and photon density as compared to the commonly used dielectric crystals, which makes them a highly desirable platform for compact photonic devices. The extreme anisotropy of the dielectric constant in these crystals is intricately linked with the anisotropic character of the phonons, which along with photon confinement leads to the rich physics of phonon polaritons. However, the chiral nature of phonons in these hyperbolic crystals have not been studied in detail. In this study, we report our observations of helicity selective Raman scattering from flakes of α-$MoO_3$. Both helicity-preserving and helicity-reversing Raman scattering are observed. Our studies reveal that helical selectivity is largely governed by the underlying crystal symmetry. This study shed light on the chiral character of the high symmetry phonons in these hyperbolic crystals. It paves the way for exploiting proposed schemes of coupling chiral phonon modes into propagating surface plasmon polaritons and realizing compact photonic circuits based on helical polarized light.

## I. INTRODUCTION

Layered van der Waal (vdW) crystals, like hexa-Boron Nitride (h-BN) [1-3] and Molybdenum trioxide [4], belong to a relatively new class of two dimensional (2D) crystals, with in-plane permittivity tensor exhibiting extreme anisotropy. Such materials demonstrate hyperbolicity i.e. the permittivity possesses both positive and negative principal parts in the infra-red spectral range [5,6]. Due to the hyperbolic nature, these 2D materials can support very high momentum photon modes with much smaller wavelengths. This makes it possible to fabricate nanometer-scale compact photonic devices [7,8]. Consequently, these materials exhibit very high photon density of states, which enables the use of Purcell enhancement in these for photonic devices [9-11].

α-$MoO_3$ has demonstrated very strong in-plane hyperbolicity [12,13], making it a strong candidate for possible nanophotonic applications, especially those requiring polarization control of photons [14-17]. The hyperbolic nature of the dielectric function in these materials is governed by the high anisotropic character of the phonons in these hyperbolic vdW materials [18]. The anisotropy in phonon modes appears due to the hybridization of longitudinal phonons with transverse phonons by quasi-static Coulomb interaction mediated by large momentum photons [3]. Strong confinement of photons in these hyperbolic vdW materials makes them an excellent platform for demonstration of polariton effects of excitons and phonons. Indeed, exciton and phonon-polariton (PhP) effects have been observed both in h-BN [1,2,19] and $MoO_3$ [4,20]. Since the optical anisotropy is governed by the Raman anisotropy, therefore the character and details of phonon mediated scattering processes in these naturally occurring hyperbolic crystals is crucial.

Phonon band structure of h-BN [21] and α-$MoO_3$ [22-24] have been studied theoretically and experimentally. However, detailed in-plane anisotropic phonon and corresponding Raman studies have only been reported recently [18,25]. Raman scattering is dominated by the phonons near the high symmetry *k*-points [18]. Polarization character of Raman photons is governed by the symmetry of the underlying crystal phonons, which provide the required energy and angular momentum for the Raman scattering processes [26]. Angular momentum must be conserved during Raman scattering of helical photons, which give rise to well-defined selection rules for helicity-based Raman scattering. These symmetry-based selection rules have been clearly observed in hexagonal 2D systems like transition metal dichalcogenides (TMDCs) [27]. The selection rules can be modified in the presence of strong electron-phonon interactions [28] or under resonant-circular excitation [29]. Indeed, helicity-based resonant Raman has proved to be an excellent tool for quantification of individual contributions of Fröhlich interaction and deformation potential to exciton-phonon interactions in TMDCs [30].

An important aspect of helicity-based Raman selection rules is the symmetry-allowed helicity reversed phonon scattering. Observation of helicity-reversed Raman scattering in 2D hexagonal TMDCs [27] have been attributed to the presence of chiral phonons at high-symmetry points in TMDCs [31]. Indeed, chiral phonons have recently been observed in $WSe_2$ (a TMDC) [32]. Chiral phonons can potentially play important role in intervalley scattering [33], phonon-driven topological states [34] and even can be a resource for quantum information processing [35].

Although, in-plane phonon anisotropy of hyperbolic materials has been recently highlighted [18,25], helicity selective Raman scattering in such systems is still unexplored. Do the phonon branches in α-$MoO_3$ support both helicity conserving and helicity reversing Raman scattering processes? Do the high-symmetry phonons in non-hexagonal crystals like α-$MoO_3$ possess chirality? Helical selectivity in hexagonal 2D crystals obeys very robust selection rules, so much so that helicity-resolved Raman scattering is a more reliable method for symmetry assignment of the mode [27]. Do non-hexagonal 2D crystals like α-$MoO_3$ demonstrate similarly robust helical selection rules for Raman scattering?

In this work, we present results of our detailed Raman investigations on flakes of α-$MoO_3$. We present in-plane Raman anisotropy of these flakes along with the polarization profile of scattered Raman modes under linear polarized excitation. More importantly, we show in-plane helicity-resolved Raman spectroscopy from flakes of α-$MoO_3$ under backscattering geometry. We observe clear helical selectivity for the Raman modes. Both helicity-conserved and helicity-switched Raman scattering processes are observed. This suggests the presence of chiral phonons in the α-$MoO_3$ crystal. Most of our results can be explained by the underlying symmetry of the crystal. Furthermore, the helicity of Raman scattered phonons is independent of number of layers for multilayered flakes.

## II. EXPERIMENTAL DETAILS

### A. Sample

Bulk α-$MoO_3$ crystals were synthesized via physical vapor deposition. Commercial $MoO_3$ powder (Sigma-Aldrich) was evaporated in a horizontal tube furnace at 785°C and bulk crystals were deposited at 560 °C. The deposition was carried out in a vacuum environment, with argon as the carrier gas for vapor transport (1 Torr). Subsequently, the bulk crystals were mechanically exfoliated using adhesive tape and thin flakes were transferred on to 300 nm $SiO_2$ on Si substrates for characterization. In this study, we measure flakes with thicknesses ranging from 150 nm to 950 nm.

### B. Spectroscopy Setup

These flakes are studied using an in-house assembled micro-Raman setup built in a confocal geometry (see Figure 2(a), 3(a) and 4(a)). The setup is fitted with a 533 nm (2.33 eV) single mode laser, which is focused on the sample via an infinity-corrected long working distance 50X objective (Olympus LMPLFLN50X). The back scattered Raman and Rayleigh scattered light is collected via the same objective. Collection of the Rayleigh scattered light is suppressed via a 533 nm notch filter (Thorlabs NF533-17). Raman spectrum is detected via an imaging spectrometer Horiba iHR550 fitted with 1200 lines/mm grating. All the measurements presented here are performed at room temperature.

### C. Experimental Results and discussion

**1. Raman modes of $MoO_3$**

α-$MoO_3$ possess octahedral symmetry with $D_{16}^{2h}$ or $Pnma$ space group [25] (see Figure 1b for a depiction of the structure). The primitive unit cell has 16 atoms, which give rise to 45 optical phonon branches and 3 acoustic phonon branches. Group theory analysis suggest 8 $A_g$, 4 $B_{1g}$, 8 $B_{2g}$, and 4 $B_{3g}$ Raman-active modes [18,25].

Molybdenum trioxide is a wide bandgap semiconductor with a direct bandgap of 3.0 eV [13,36] and a monolayer thickness of 0.7 nm [37]. The Raman studies reported here are based on laser scattering of energy of 2.33 eV, i. e. below the bandgap. The flakes are oriented such that the crystallographic *c*- and *a*-axes are aligned in-plane (see Fig. 1(b) for reference). The laser is incident at normal angle, therefore, the Raman configuration for these measurements can be represented by $-y(e_i, e_s)y$, where $-y$ and $y$ represents the direction of the incident and scattering laser in the lab frame, while $e_i$ and $e_s$ represent the *in-plane* polarization state of the incident and scattered light, respectively. Figure 1a shows a typical Raman spectrum from an $MoO_3$ flake, in which 8 distinct Raman modes are readily observed. These modes include 5 $A_g$ modes at 336, 364, 482, 817 and 992 cm$^{-1}$. The mode at 283 cm$^{-1}$ has $B_{2g}$ character, while the mode at 666 cm$^{-1}$ has a $B_{3g}$ character. These designations are based on previously reported works [18,28,38,39] and helical selection rules for Raman scattering as discussed below. The peak at 519 cm$^{-1}$ corresponds to the Silicon substrate. Although, the theory suggests 24 Raman modes [18], however, we only observe 8 modes. Detection of 11 low-energy Raman modes is blocked by the notch filter used in this study, while the other modes are not visible due to the degeneracies between various modes [18].

***************************************************************************
********************************  Figure 1  ********************
***************************************************************************

**2. Raman Anisotropy of $MoO_3$**

The in-plane hyperbolic optical response of flakes of $MoO_3$ is directly related to the in-plane phonon anisotropy. We probe the in-plane Raman anisotropy by analyzing the Raman mode intensity as a function of the relative angle between the incident laser polarization and crystallographic *c*-axis. We performed a series of anisotropy measurements and compare our results with the previous studies [18,25]. The Raman scattering cross-section of a specific mode directly depends on the $I \propto |e_s \cdot R \cdot e_i|^2$, where $e_i$ and $e_s$ denote the polarization state of the incident laser and scattered light. The Raman susceptibility is represented by the Raman tensors, represented here by symbol $R$. Each element of the tensor $R_{ij}(\gamma)$ represents the corresponding susceptibility of the $\gamma$-th Raman active mode in the crystal coordinates near the Brillouin zone boundary. The tensors used in this study are adapted from Reference [18], which derives these tensors using group theory treatment of the polarizability tensor. The polarizability tensor is calculated using density functional theory (DFT). These tensors for the Raman active modes of $MoO_3$ are:

$$R(A_g) = \begin{pmatrix} a & 0 & 0 \\ 0 & b & 0 \\ 0 & 0 & c \end{pmatrix} \qquad R(B_{1g}) = \begin{pmatrix} 0 & d & 0 \\ d & 0 & 0 \\ 0 & 0 & 0 \end{pmatrix} \qquad R(B_{2g}) = \begin{pmatrix} 0 & 0 & e \\ 0 & 0 & 0 \\ e & 0 & 0 \end{pmatrix} \qquad R(B_{3g}) = \begin{pmatrix} 0 & 0 & 0 \\ 0 & 0 & f \\ 0 & f & 0 \end{pmatrix} \quad (1)$$

The Raman scattering cross section can then be represented in the lab frame as $I \propto |e_s \cdot t^T \cdot R \cdot t \cdot e_i|^2$, where a transformation matrix $t$ converts the crystal coordinates into the lab frame.

$$t = \begin{pmatrix} \cos\theta & 0 & -\sin\theta \\ 0 & 1 & 0 \\ \sin\theta & 0 & \cos\theta \end{pmatrix} \qquad (2)$$

The angle $\theta$ is the angle between the crystallographic $c$ axis and $z$ axis of the lab frame. The crystallographic $b$ axis coincides with the $y$ axis of the lab frame. The intensities of the Raman active modes can then be written as

$$I(A_g) \propto \left| \begin{pmatrix} 1 \\ 0 \\ 0 \end{pmatrix}^T \begin{pmatrix} \cos\theta & 0 & -\sin\theta \\ 0 & 1 & 0 \\ \sin\theta & 0 & \cos\theta \end{pmatrix}^T R(A_g) \begin{pmatrix} \cos\theta & 0 & -\sin\theta \\ 0 & 1 & 0 \\ \sin\theta & 0 & \cos\theta \end{pmatrix} \begin{pmatrix} 1 \\ 0 \\ 0 \end{pmatrix} \right|^2 \qquad (3)$$

$$I(A_g) \propto |c(\cos\theta)^2 + a(\sin\theta)^2|^2 \qquad (4)$$

Similarly,

$$I(B_{2g}) \propto |e \sin 2\theta|^2 \qquad (5)$$

$$I(B_{1g}) = I(B_{3g}) = 0 \qquad (6)$$

For Raman anisotropy measurements (see Figure 2(a), the sample is fixed while the incident laser polarization is rotated with respect to the in-plane crystallographic *c* axis using a $\lambda/2$ retarder in the incident beam of the setup (see Figure 2(b)). The detection polarization angle is selected via a set of $\lambda/2$ retarder and a linear polarizer in the detection path. The detection polarization is kept parallel to the incident polarization for all measurements here. Raman spectra are taken at varied angle $\theta$ between the incident polarization and the crystal axis *c* (Fig. 2(b)). The resulting series of Raman spectra are depicted as color plots in Fig. 2(c), which shows clear dependence of Raman modes signatures on $\theta$. Polar plots of the normalized Raman signal as a function of $\theta$ are shown in Fig. 2(d)-(h) for 5 different modes.

Equations (4)-(6) suggest that only the $A_g$ and $B_{2g}$ modes should be observable. However, we do observe weak signal from the $B_{3g}$ mode at 666 cm$^{-1}$. The polar plots of $A_g$ and$B_{2g}$ modes clearly show the in-plane anisotropy. Polarization of the $A_g$ modes at 817 and 992 cm$^{-1}$ are predominantly aligned along the crystallographic *c*-axis (i. e. $|c| > |a|$). The $A_g$ mode at 364 cm$^{-1}$ is strongly polarized along the *a*-axis ($|c| \ll |a|$). These observations agree with the previously reported Raman anisotropy studies [18,25]. However, unlike these studies, the $A_g$ mode at 336 cm$^{-1}$ has maxima both along 0° and 90°. Fitting using Eq. (4) implies $|c| \sim |a|$. For the $B_{2g}$ mode at 283 cm$^{-1}$, the maxima of the intensity occurs at +45°, +135° and so on. Unlike that of $A_g$ mode, the polarization of $B_{2g}$ is aligned along the diagonal of the *a*-*c* plane.

*****************************************************************************
******************************** Figure 2 ********************
*****************************************************************************

**3. Polarization-resolved Raman**

To further emphasize the difference in the Raman scattering of $A_g$ and $B_{2g}$ modes, we demonstrate polarization-resolved Raman scattering under constant linear excitation. The linear polarization of incident laser is aligned along the in-plane crystal *c*-axis using a $\lambda/2$ retarder in the incident beam (see Figure 3(a)). Raman scattering is detected as a function of angle ($\phi$) between the detection polarization and the crystal *c*-axis (see Fig. 3(b)). The angle $\phi$ is gradually varied using a set of $\lambda/2$ retarder and a linear polarizer in the detection path. Figure 3(c) shows color plot of the Raman signal as a function of detection polarization angle $\phi$.

Normalized Raman mode intensities extracted from these measurements are depicted in Fig. 3(d) as polar plots for three $A_g$ modes (at 336, 817 and 992 $cm^{-1}$) and a $B_{2g}$ mode (at 283 $cm^{-1}$). These polar plots reveal the polarization state of backscattered Raman modes for the case when the incident laser polarization is aligned along crystallographic $c$-axis. All the scattered Raman modes are linearly polarized, however, the $A_g$ modes have polarization predominantly along the *c*-axis, while the $B_{2g}$ mode has polarization along the diagonal of the *a*-*c* plane. This implies a $\pi/4$ rotation of polarization of the incident photon during the Raman scattering process, further highlighting the difference in Raman scattering between $A_g$ and $B_{2g}$ modes.

*****************************************************************************
********************************       Figure 3       ********************
*****************************************************************************

## 4. Helicity-resolved Raman

Recent studies on two dimensional semiconductors have clearly revealed robust selection rules for helicity-resolved Raman scattering [28]. Interestingly, apart from momentum conserving helical Raman scattering, clear signatures of helically switched Raman scattering are reported [28]. Such large momentum transfer was initially attributed to mixing of $s$- and $p$-orbitals in the conduction band at high symmetry points along with the wave vector dependent Berry curvature. Later studies indicated the chiral nature of the phonons at high symmetry point in hexagonal 2D vdW crystals as the dominant mechanism responsible for helicity switching [27] . However, such helical selectivity of the Raman modes in biaxial hyperbolic crystals like α-$MoO_3$ has yet to be reported.

The experimental setup for helicity-resolved Raman scattering measurements is depicted in Figure 4(a). Laser of a specific helicity is prepared using a linear polarizer and an achromatic $\lambda/4$ retarder (from Bernhard Halle Nachfl. GmbH). Scattered Raman signal from the sample is collected by the confocal objective. The same $\lambda/4$ retarder then projects helical polarizations into a polarization plane for propagation in the detection path, such that the plane is defined by co- and cross-polarized polarization components. The co-polarized component in the detection refers to the detection of Raman signal with the same helicity as that of the incident laser, while the cross-polarized component represents the detection of opposite helical polarization. A set of $\lambda/2$ retarder and linear polarizer in the detection path is used for helicity selective detection.

Considering the incident laser along the $y$ axis (lab frame as well as the crystal coordinates), the scattered intensities for the Raman active modes under σ+ incident laser is given as:

$$I(A_g)_{co} \propto \left| \begin{pmatrix} 1 \\ 0 \\ i \end{pmatrix}^T R(A_g) \begin{pmatrix} 1 \\ 0 \\ i \end{pmatrix} \right|^2 = |a+c|^2 \qquad (7)$$

$$I(A_g)_{cross} \propto \left| \begin{pmatrix} 1 \\ 0 \\ -i \end{pmatrix}^T R(A_g) \begin{pmatrix} 1 \\ 0 \\ i \end{pmatrix} \right|^2 = |a-c|^2 \qquad (8)$$

$$I(B_{2g})_{co} \propto \left| \begin{pmatrix} 1 \\ 0 \\ i \end{pmatrix}^T R(B_{2g}) \begin{pmatrix} 1 \\ 0 \\ i \end{pmatrix} \right|^2 = 0 \qquad (9)$$

$$I(B_{2g})_{cross} \propto \left| \begin{pmatrix} 1 \\ 0 \\ -i \end{pmatrix}^T R(B_{2g}) \begin{pmatrix} 1 \\ 0 \\ i \end{pmatrix} \right|^2 = 4|e|^2 \qquad (10)$$

$$I(B_{1g})_{cross} = I(B_{1g})_{co} = 0 = I(B_{3g})_{cross} = I(B_{3g})_{co} \qquad (11)$$

Using these calculations, only $A_g$ and $B_{2g}$ modes should be observed under helical excitation. The $A_g$ mode has weaker selection rule for scattering as compared to $B_{2g}$, which should strictly have cross-polarized character. Raman scattering of $B_{1g}$ and $B_{3g}$ is forbidden under helical scattering. Remember these selection rules are only valid for the case when incident laser is perpendicular on an *a*-*c* crystallographic plane of $MoO_3$ flake.

Figure 4 shows helicity-resolved Raman spectra of α-$MoO_3$ flake (of thickness 150 nm) under σ+ excitation. Almost complete switching off of the Raman mode in the co-polarized spectrum (σ+ detection: blue line) and the cross-polarized (σ- detection: orange line) is an indicator of the helical selectivity of the Raman modes. The $A_g$ modes at 336, 817and 992 cm$^{-1}$ are dominantly co-polarized, however the degree of helical selectivity is higher for the mode at 336 cm$^{-1}$ (vanishingly small $|a-c|^2$) and low for the mode at 817 cm$^{-1}$ ($|a-c|^2 > 0$). The $B_{2g}$ at 283 cm$^{-1}$ as expected has vanishing co-polarized intensity as compared to the intensity of backscattered cross-polarized component, as predicted from the Raman tensor. However, the $B_{3g}$ mode at 666 cm$^{-1}$ has a non-vanishing intensity under helical excitation, contrary to the predictions. It shows a dominant cross-polarized character as well. We attribute observation of this forbidden Raman mode to the non-normal part of the incident beam as the laser is focused onto the flake surface.

These measurements are repeated and verified on several flakes of α-$MoO_3$ with thicknesses ranging from 150 nm to 950 nm (check section 3 of supplementary material). The helically selective scattering of these Raman modes is unchanged under σ- excitation as well (see Fig. S3 in supplementary material).

*************************************************************************

******************************* Figure 4 *******************
***************************************************************************

The backscattered Raman signal from the flake is detected through a quarter waveplate (QWP), which project the helicities into a linear polarization plane, whose axes are defined by the co- and cross-polarized helical components as shown in Fig. 5(b). To further explore the helical selectivity of the Raman modes, we detect Raman signal at varying polarization angles within the co- and cross-polarized plane. The resulting evolution of the Raman signal is shown in Fig. 5(a). Polar plots of the normalized Raman signal extracted from this series are depicted in Figure 5(c)-(g).

All the $A_g$ modes have higher intensities when detected along the co-polarized axis. The cross-polarized component for the 336 cm$^{-1}$ $A_g$ mode is ~12% of the maximum intensity, while that of 992 cm$^{-1}$ is 20% and that of 817 cm$^{-1}$ mode has a large cross-polarized component of 27%. Both the $B_{2g}$ and $B_{3g}$ modes are predominantly cross-polarized, which implies that the crystal imparts an angular momentum to the photon thereby switching the helicity of the scattered photons. $B_{2g}$ has a higher degree of polarization reversal (less than 1% of co-polarized scattering) as compared to the $B_{3g}$ mode (~8% of co-polarized Raman scattering). Scattering of $B_{2g}$ Raman mode involves almost complete reversal of the helicity of the incident photon. These results point toward chiral character of the orthorhombic α-$MoO_3$ crystal, which requires further detailed theoretical explanation.
***************************************************************************
******************************* Figure 5 *******************
***************************************************************************

Helical selectivity of Raman mode in hexagonal 2D crystals has been used as a robust indicator for the symmetry designation of a Raman mode [27]. Equations (9)-(11) shows that $B_{2g}$ and $B_{3g}$ Raman modes of α-$MoO_3$ have clear helical selectively. This selectivity can be used for reliable mode identification as well as crystal plane identification. For example, the Raman mode at 283 cm$^{-1}$ has been assigned as a $B_{2g}$ mode in Ref. [25], while it is indicated as $B_{1g}$ in Ref. [18]. Helicity dependent Raman scattering rules forbids scattering of $B_{3g}$ modes while it allows cross-polarized scattering of $B_{2g}$ mode. Since the 283 cm$^{-1}$ in Fig. 4 has cross-polarized character and has clearly higher intensity than the helicity forbidden 666 cm$^{-1}$, we can assign it as a $B_{2g}$ mode. Relatively weaker intensity of $B_{3g}$ is even more evident in Figs. S2 and S3 of supplementary material.

The helicity selection rules clearly depend on the orientation of the flake. If we assume that the plane of the flake is defined by crystallographic *a-c* axis, then only the $B_{2g}$ mode should have cross-polarized scattering, while scattering of $B_{1g}$ and $B_{3g}$ should be forbidden. However, if we assume the plane to be made of *b-c* crystallographic axes, then scattering of $B_{1g}$ and $B_{2g}$ becomes forbidden and cross-polarized component of $B_{2g}$ should be non-zero. Figure 4 and Figs. S2 and S3 of the supplementary material show that $B_{3g}$ has considerably smaller intensity than $B_{2g}$, hence we infer that the surface of the $MoO_3$ flakes in our studies is defined by *a*-*c* crystal plane (see section 4 of supplementary material for detailed calculations).

## III. CONCLUSION

In conclusion, we observe the presence of clear helicity selection rules in Raman scattering from $MoO_3$ flakes. Almost complete helicity switching under Raman scattering points to strong phonon chirality around the high symmetry points in Brillouin zone of $MoO_3$ crystal. The helical selectivity largely agrees with the symmetry-based Raman tensors, however, a detailed theoretical explanation of the chirality of phonon bands in α-$MoO_3$ is needed.

The hyperbolic nature of flakes of α-$MoO_3$ makes them ideal candidates for nano-photonics applications apart from the highly rich physics of in-plane anisotropic phonon polariton. The chiral nature of the highly anisotropic phonons in this material system can play a crucial role in proposals which combine this hyperbolicity with spin-orbit coupling resulting in novel surface plasmon modes [40]. Especially in schemes which rely on coupling of chiral Raman modes into propagating surface plasmon polaritons [41]. Understanding of the chiral nature of the phononic bands can be vital for photonic circuits involving helical metamaterials [42,43].

## Acknowledgements

S. A. A. acknowledges PhD funding from SBASSE LUMS. A. U. acknowledge funding from LUMS via FIF-523.

## Data Availability

The data that support the findings of this study are available from the corresponding author upon reasonable request.

## References

[1] Caldwell, J. D., Kretinin, A. V., Chen, Y., *et al*., “Sub-diffractional volume-confined polaritons in the natural hyperbolic material hexagonal boron nitride” Nature communications, **5**, 1-9 (2014).

[2] Dai, S., Quan, J., Hu, G. et *al*., “Hyperbolic phonon polaritons in suspended hexagonal boron nitride”. Nano letters **19**, 1009-1014 (2018).

[3] Low, T., Chaves, A., Caldwell, J. D., *et al*., “Polaritons in layered two-dimensional materials”, Nature materials, **16**, 182-194, (2017)

[4] Ma, W., Alonso-González, P., Li, S., *et al*. “In-plane anisotropic and ultra-low loss polaritons in a natural van der Waals crystal”. Nature **562**, 557-562 (2018).

[5] Poddubny, A., Iorsh, I., Belov, P., and Kivshar, Y., “Hyperbolic metamaterials”. Nature photonics, **7**, 948-957 (2013).

[6] Guo, Z., Haitao J., and Hong C. "Hyperbolic metamaterials: From dispersion manipulation to applications." Journal of Applied Physics **127**, 071101 (2020).

[7] Zhukovsky, S.V., Andryieuski, A., Sipe, J. E., *et al*. "From surface to volume plasmons in hyperbolic metamaterials: general existence conditions for bulk high-k waves in metal-dielectric and graphene-dielectric multilayers." Physical Review B **90**, 155429 (2014).

[8] Takayama, O. and Lavrinenko, A.V., "Optics with hyperbolic materials." JOSA B **36**, F38-F48 (2019).

[9] Noginov, M. A., H. Li, Yu A. Barnakov, D., *et al*. "Controlling spontaneous emission with metamaterials." Optics letters **35**, 1863-1865 (2010).

[10] Tumkur, T., Zhu, G., Black, P., *et al*. "Control of spontaneous emission in a volume of functionalized hyperbolic metamaterial." Applied Physics Letters **99**, 151115 (2011).

[11] Krishnamoorthy, H.N., Jacob, Z., Narimanov, E., *et al*. "Topological transitions in metamaterials." Science **336**, 205-209 (2012).

[12] Zheng, Z., Xu, N., Oscurato, S.L., *et al*. "A mid-infrared biaxial hyperbolic van der Waals crystal." Science advances **5**, eaav8690 (2019).

[13] Balendhran, S., Walia, S., Nili, H., *et al*. "Two-dimensional molybdenum trioxide and dichalcogenides." Advanced Functional Materials **23**, 3952-3970 (2013).

[14] Folland, Thomas G., and Joshua D. Caldwell. "Precise control of infrared polarization using crystal vibrations." Nature **562**, 499-501 (2018).

[15] Dixit, S., Sahoo, N.R., Mall, A., *et al*., "Mid infrared polarization engineering via sub-wavelength biaxial hyperbolic van der Waal crystals", Scientific Reports **11**, 6612 (2021)

[16] Li, P., Dolado, I., Alfaro-Mozaz, F.J., *et al*. "Infrared hyperbolic metasurface based on nanostructured van der Waals materials." Science **359**, 892-896 (2018).

[17] Nicholls, L.H., Rodríguez-Fortuño, F.J., Nasir, M.E., *et al*. "Ultrafast synthesis and switching of light polarization in nonlinear anisotropic metamaterials." Nature Photonics **11**, 628-633 (2017).

[18] Wen, M., Chen, X., Zheng, Z., *et al*. "In-Plane Anisotropic Raman Spectroscopy of van der Waals α-$MoO_3$." The Journal of Physical Chemistry *C* **125**, 765-773 (2020).

[19] Woessner, A., Lundeberg, M.B., Gao, Y., *et al*. "Highly confined low-loss plasmons in graphene–boron nitride heterostructures". Nature materials **14**, 421-425 (2015).

[20] Zheng, Z., Chen, J., Wang, Y., *et al*., "Highly confined and tunable hyperbolic phonon polaritons in van der Waals semiconducting transition metal oxides", Advanced Material **30**, 1705318 (2018).

[21] Michel, K. H., and Verberck, B. "Phonon dispersions and piezoelectricity in bulk and multilayers of hexagonal boron nitride." Physical Review B **83**, 115328 (2011).

[22] Lee, S.H., Seong, M.J., Tracy, C.E., *et al*., "Raman spectroscopic studies of electrochromic α-$MoO_3$ thin films". Solid State Ionics **147**, 129-133 (2002).

[23] Camacho-López, M.A., Escobar-Alarcón, L., Picquart, M., *et al*., "Micro-Raman study of the m-$MoO_2$ to α-$MoO_3$ transformation induced by cw-laser irradiation" Optical Materials **33**, 480-484 (2011).

[24] Spevack, P. A., and McIntyre, N. S. "Thermal reduction of molybdenum trioxide", The Journal of Physical Chemistry **96**, 9029-9035 (1992).

[25] Zheng, B., Wang, Z., Chen, Y., *et al*. "Centimeter-sized 2D α-$MoO_3$ single crystal: growth, Raman anisotropy, and optoelectronic properties." 2D Materials **5**, 045011 (2018).

[26] Powell, Richard C., "Symmetry, group theory, and the physical properties of crystals" Vol. **824**. New York: Springer, 2010.

[27] Chen, S. Y., Zheng, C., Fuhrer, M. S., & Yan, J. "Helicity-resolved Raman scattering of $MoS_2$, $MoSe_2$, $WS_2$, and $WSe_2$ atomic layers.", Nano letters **15**, 2526-2532 (2015).

[28] Zhao, Y., Han, S., Zhang, J., *et al*. "Helicity-resolved resonant Raman spectroscopy of layered $WS_2$." Journal of Raman Spectroscopy **52**, 525-531 (2021).

[29] Huang, J., Liu, Z., Yang, T. and Zhang, Z. "New selection rule of resonant Raman scattering in $MoS_2$ monolayer under circular polarization." Journal of Materials Science & Technology **102**, 132-136 (2022).

[30] Zhao, Y., Zhang, S., Shi, Y., *et al*. "Characterization of Excitonic Nature in Raman Spectra Using Circularly Polarized Light." ACS nano **14**, 10527-10535 (2020).

[31] Zhang, L., and Qian, N. "Chiral phonons at high-symmetry points in monolayer hexagonal lattices", Physical review letters **115**, 115502 (2015).

[32] Hanyu Z., Jun Y., Ming-Yang L. et al., "Observation of chiral phonons" Science **359**, 579–582 (2018).

[33] Carvalho, B.R., Wang, Y., Mignuzzi, S., *et al*. "Intervalley scattering by acoustic phonons in two-dimensional $MoS_2$ revealed by double-resonance Raman spectroscopy." Nature communications **8**, 1-8 (2017).

[34] Forst, M., Roman Mankowsky, and Andrea Cavalleri. "Mode-selective control of the crystal lattice." Accounts of chemical research **48**, 380-387 (2015).

[35] Lee, K. C., Sprague, M. R., Sussman, B. J., *et al*. "Entangling macroscopic diamonds at room temperature." Science **334**, 1253-1256 (2011).

[36] Vos, M.F., Macco, B., Thissen, N.F., *et al*. "Atomic layer deposition of molybdenum oxide from $(N\underline{t}Bu)_2(NMe_2)_2Mo$ and $O_2$ plasma." Journal of Vacuum Science & Technology A: Vacuum, Surfaces, and Films **34**, 01A103 (2016).

[37] Cai, L., McClellan, C.J., Koh, A.L., *et al*. "Rapid flame synthesis of atomically thin $MoO_3$ down to monolayer thickness for effective hole doping of $WSe_2$." Nano letters **17**, 3854-3861 (2017).

[38] Syrbu, N. N., and I. G. Stamov. "The superposition of the lattice radiation and reflectivity spectra of $MoO_3$ and $PbMoO_4$ crystals." Crystal Research and Technology **29**, 133-148 (1994).

[39] Eda, Kazuo. "Raman spectra of hydrogen molybdenum bronze, $H_{0.30}MoO_3$." Journal of Solid-State Chemistry **98**, 350-357 (1992).

[40] Nemilentsau, A., Stauber, T., Gómez-Santos, G., *et al*. "Switchable and unidirectional plasmonic beacons in hyperbolic two-dimensional materials." Physical Review B **99**, 201405 (2019).

[41] Guo, Q., Fu, T., Tang, J., *et al*. "Routing a chiral Raman signal based on spin-orbit interaction of light." Physical review letters **123**, 183903 (2019).

[42] Kilic, U., Hilfiker, M., Ruder, A., *et al*. "Broadband enhanced chirality with tunable response in hybrid plasmonic helical metamaterials." Advanced Functional Materials **31**, 2010329 (2021).

[43] Kaschke, Johannes, and Martin Wegener. "Optical and infrared helical metamaterials." Nanophotonics **5**, 510-523 (2016).

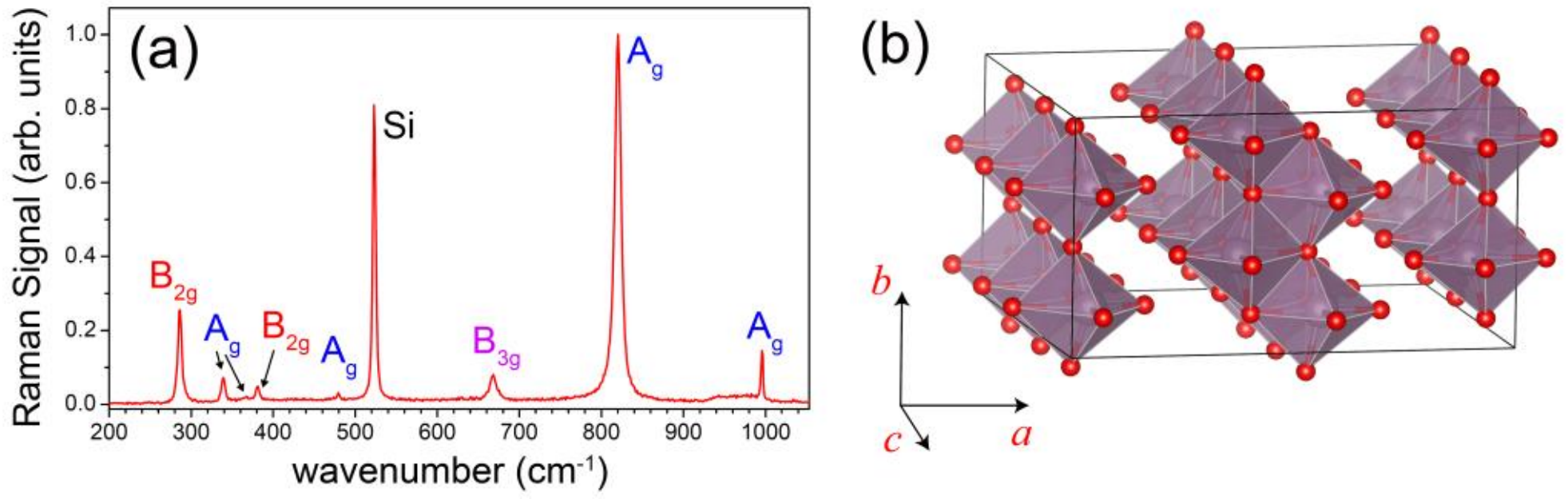

Figure 1: (a) Raman spectrum obtained via scattering of 2.33 eV laser from $MoO_3$ flake. Excitation power is kept at 1.5 mW and the temperature at 25 $C°$. The symmetry-based identification of each peak is indicated. (b) Crystal structure of $MoO_3$. The red spheres represent O atoms while the Mo atoms are enclosed at the center of the octahedrons. Plane surfaces define the orthorhombic structures. Crystallographic axes are shown.

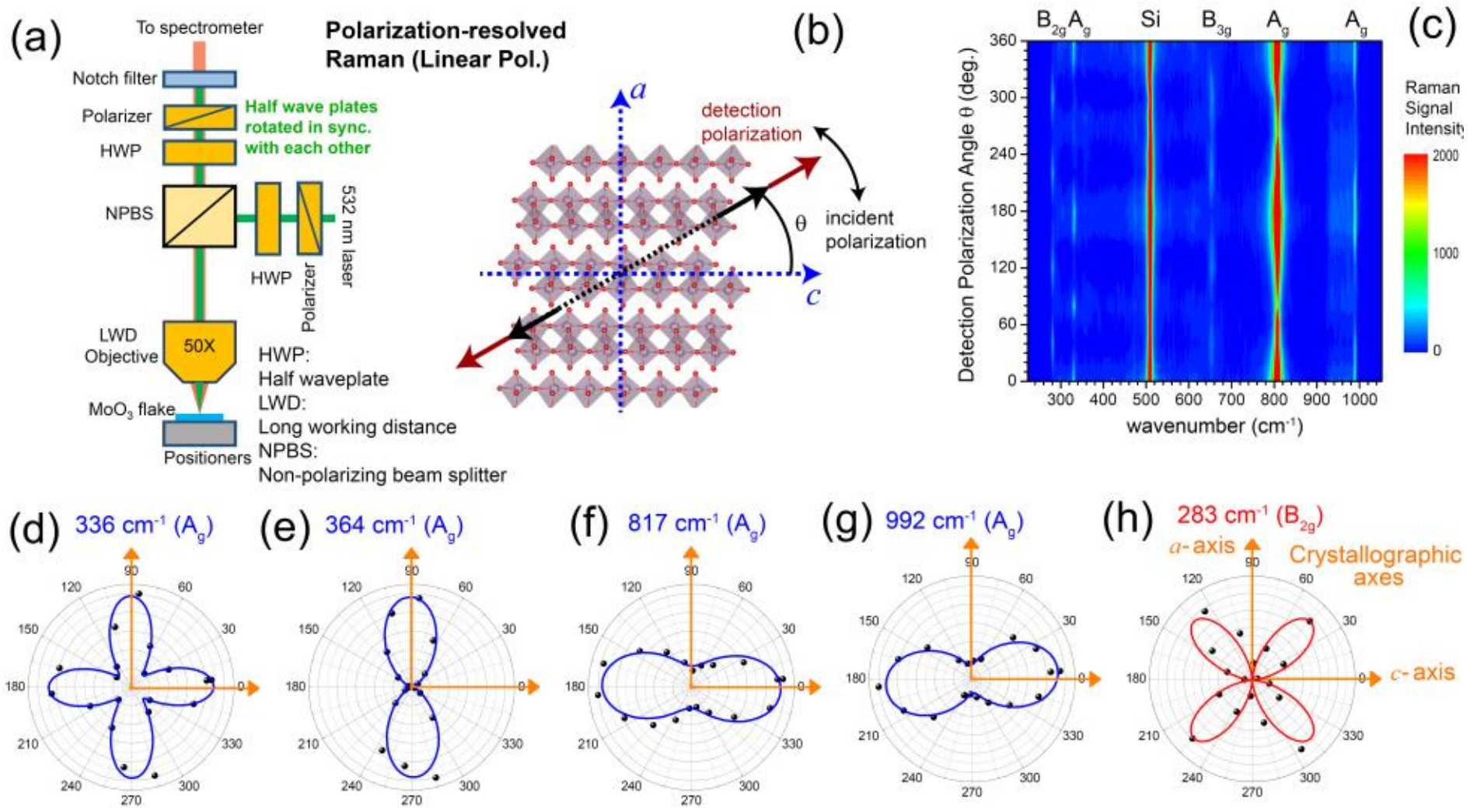

Figure 2: (a) Experimental setup for probing Raman anisotropy. A half waveplate (HWP) and a polarizer in the incident path defines the incident polarization. Another HWP and linear polarizer select detection of polarization component parallel to incident polarization. The flake is stationery as the in-plane polarization of incident laser is varied. (b) Varying incident polarization is represented by black dashed line, while the detection polarization is represented by dark red line. The crystal plane defined by *b*-*c* axes is shown. These crystallographic axes stay stationery during this experiment, while incident and detection polarization is rotated gradually, which changes $\theta$, the angle between laser polarization and crystal *c*-axis. The setup can be represented as $-x(\cos\theta\,, \sin\theta)x$. (c) Color plot of Raman modes intensity as a function of angle $\theta$. (d)-(e) Polar plots of normalized Raman intensity as extracted from the series of spectra shown in (c). Central wavenumber of each Raman mode is indicated along with its group representation. Polar plots of 4 $A_g$ modes and one $B_{1g}$ mode are shown here. Black points represent experimental data while the solid blue and red line are theoretical fit to the data using equations (4)-(6). Fitted parameters are tabulated in Table S2 of supplementary material.

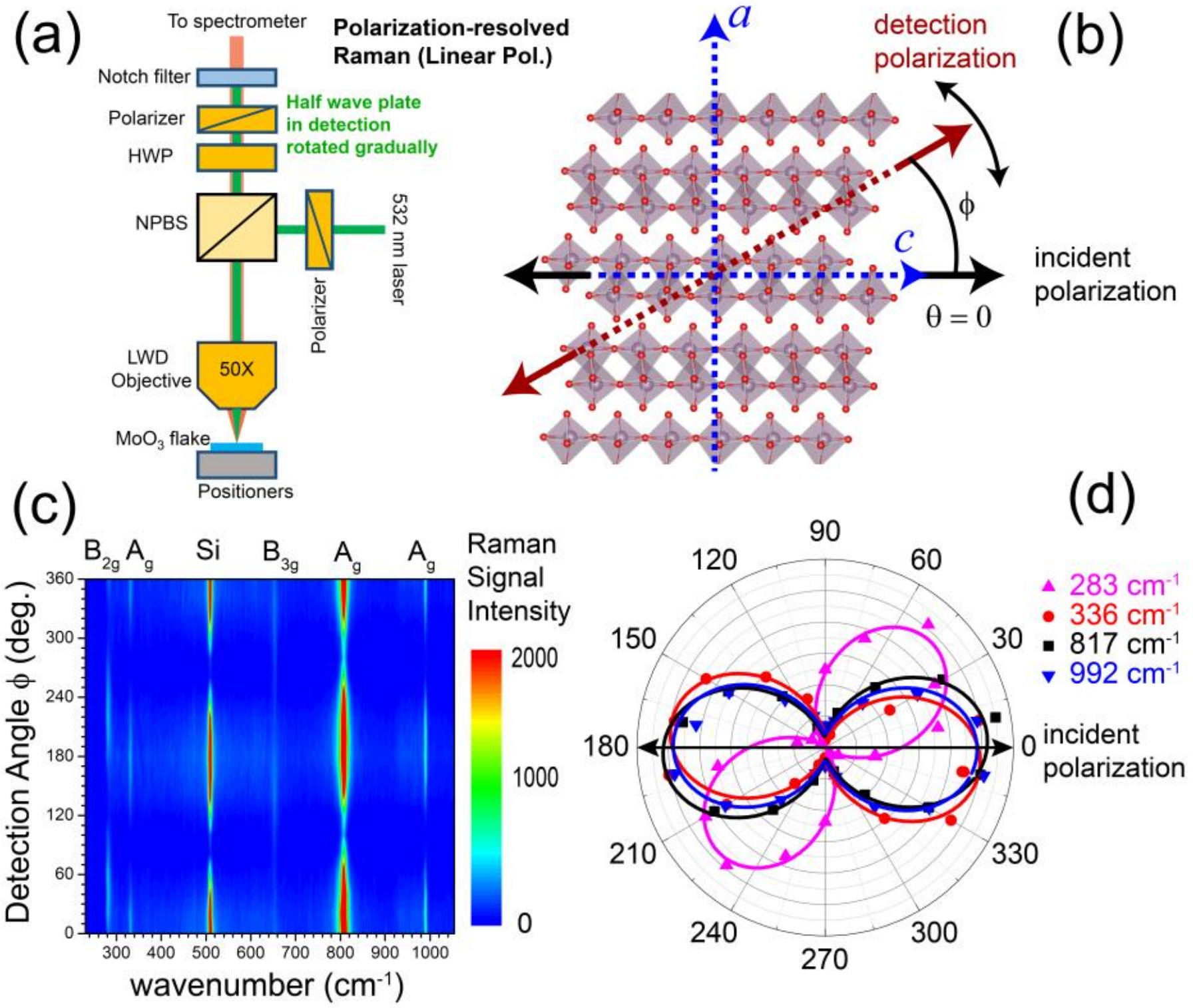

Figure 3: (a) Experimental setup for polarization-resolved measurement. The incident polarization is fixed along a crystallographic axis (*c*-axis) using a polarizer in the incidence part. The detection polarization angle is varied via a HWP and a linear polarizer in the detection path. (b) The sample stays stationery during the experiment i. e. the angle between incident polarization and crystal *c*-axis is $\theta = 0$. The detection (represented by dark red dashed line) is gradually varied i.e. Raman spectral are taken at various angles $\phi$, which represent the angle between the crystal *c*-axis and detection polarization. (c) Color plot which shows the evolution of Raman signatures as a function of $\phi$. (d) Polar plots of normalized Raman intensities extracted from the series of spectra shown in (c). Polar plot for $A_g$ modes at 336 cm$^{-1}$ (red), 817 cm$^{-1}$ (black) and 992 cm$^{-1}$ (blue) and the $B_{2g}$ mode at 283 cm$^{-1}$ (magenta). Points represent experimental data while the solid lines are fit to the data.

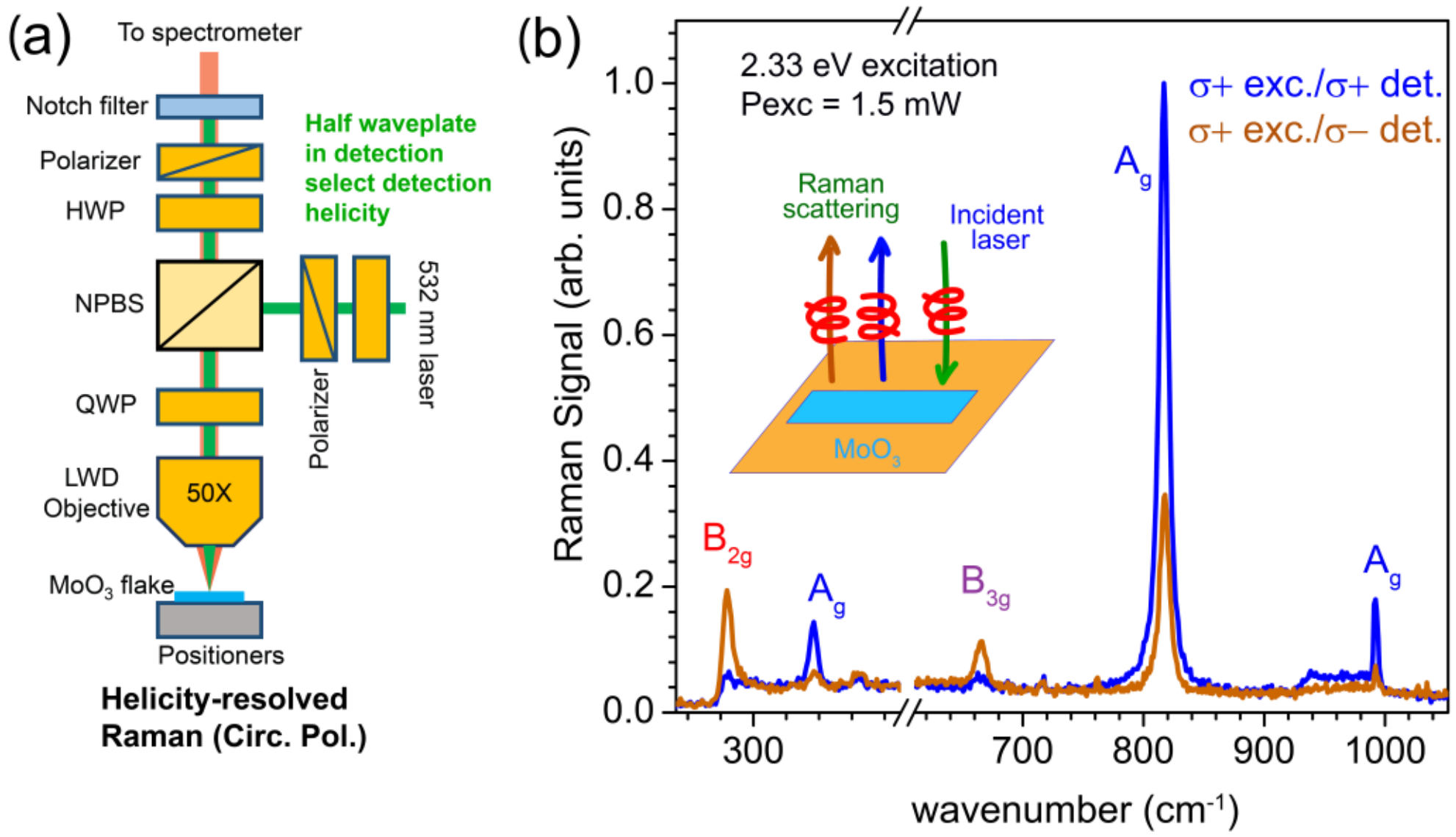

Figure 4: (a) Experimental setup for helicity-resolved Raman studies on flakes of $MoO_3$. Linear polarization for incident beam is defined by a linear polarizer. Linear polarization is converted to circular polarization of a specific helicity using a quarter waveplate (QWP) just below the beam splitter. The same QWP project the helicities in the backscattered Raman signal into a polarization plane defined by co- and cross-polarized components. Specific component is selected using HWP in the detection path. (b) QWP is configured such that σ+ polarized laser is incident on the flake. The HWP in the detection then select the co-polarized (σ+) component in the detection path. The corresponding Raman spectrum is shown in blue color. The HWP then select the cross-polarized (σ-) component for the detection. The Raman spectrum for this configuration is plotted in orange. The experimental configuration can be represented as $-y(\sigma+,\sigma+)y$ for the co-polarized detection while $-y(\sigma+,\sigma-)y$ for the cross-polarized configuration as schematically depicted in the inset. The Raman peak from Silicon is not shown, for clear visualization of the $MoO_3$ Raman modes.

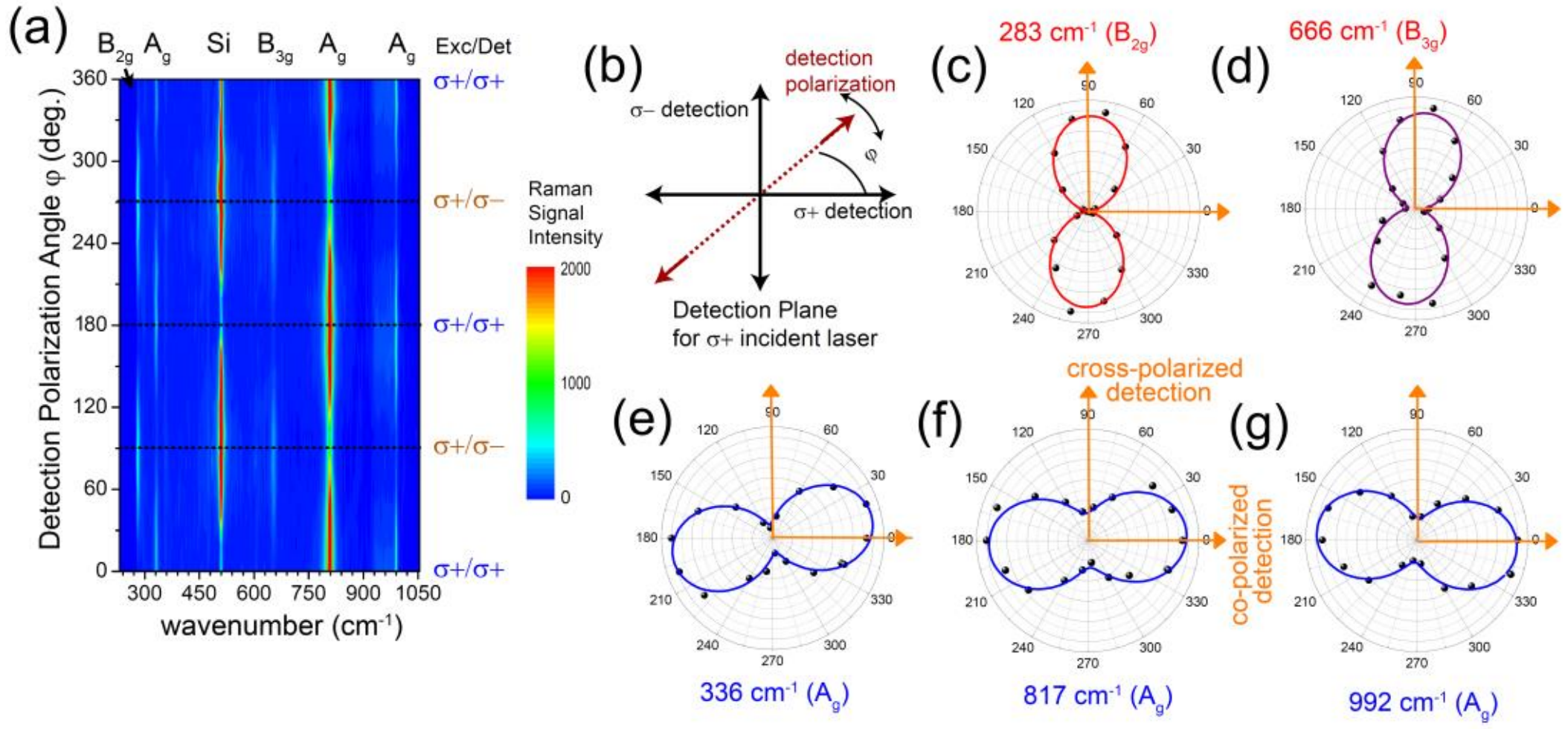

Figure 5: (a) Color plot of the Raman signal as a function of the angle φ between detection polarization and the co-polarized axis in the detection plane defined by the co- and cross-polarized components of backscattered signal. (b) Shows the schematic of this detection plane and the orientation of the detection polarization for a specific value of φ. (c)-(g) Polar plots of the normalized Raman mode intensity as a function of φ. Black dots represent experimental results while solid lines represent the fits extracted from the corresponding Raman tensor. Blue solid line represents fitting of $A_g$ mode, red solid line is the fit to the $B_{2g}$ mode, while purple line fit the intensity of the $B_{3g}$ mode. The orange lines show the detection polarization angle for which the detected Rayleigh scattering is maximum (co-polarized) or minimum (cross-polarized).

# Supplementary Material: Helicity-selective Raman scattering from in-plane anisotropic α-$MoO_3$

The supplementary material with this manuscript includes 5 sections. Section 1 discuss and show thickness determination of a few flakes of $MoO_3$ used in this study. Section 2 is detailed description of how polarization angles in the detection are determined for the three different types of angle-resolved measurements. Section 3 depicts helicity-resolved Raman spectra for 6 different $MoO_3$ flakes both under $\sigma$+ and $\sigma$- excitations. Section 4 discusses the use of helicity resolved Raman for determination of sample orientation. Section 5 of supplementary material tabulate the Raman tensor elements extracted using the fitting of experimental data.

## 1. Atomic Force Microscopy Scan of $MoO_3$ flakes

Atomic force microscopy (AFM) scans of the typical $MoO_3$ flakes reported in this study are shown in Figure S1. These flakes have thicknesses of 140, 200 and 300 nm, which implies that in this study we are not dealing with a few layer flakes, but rather multilayered.

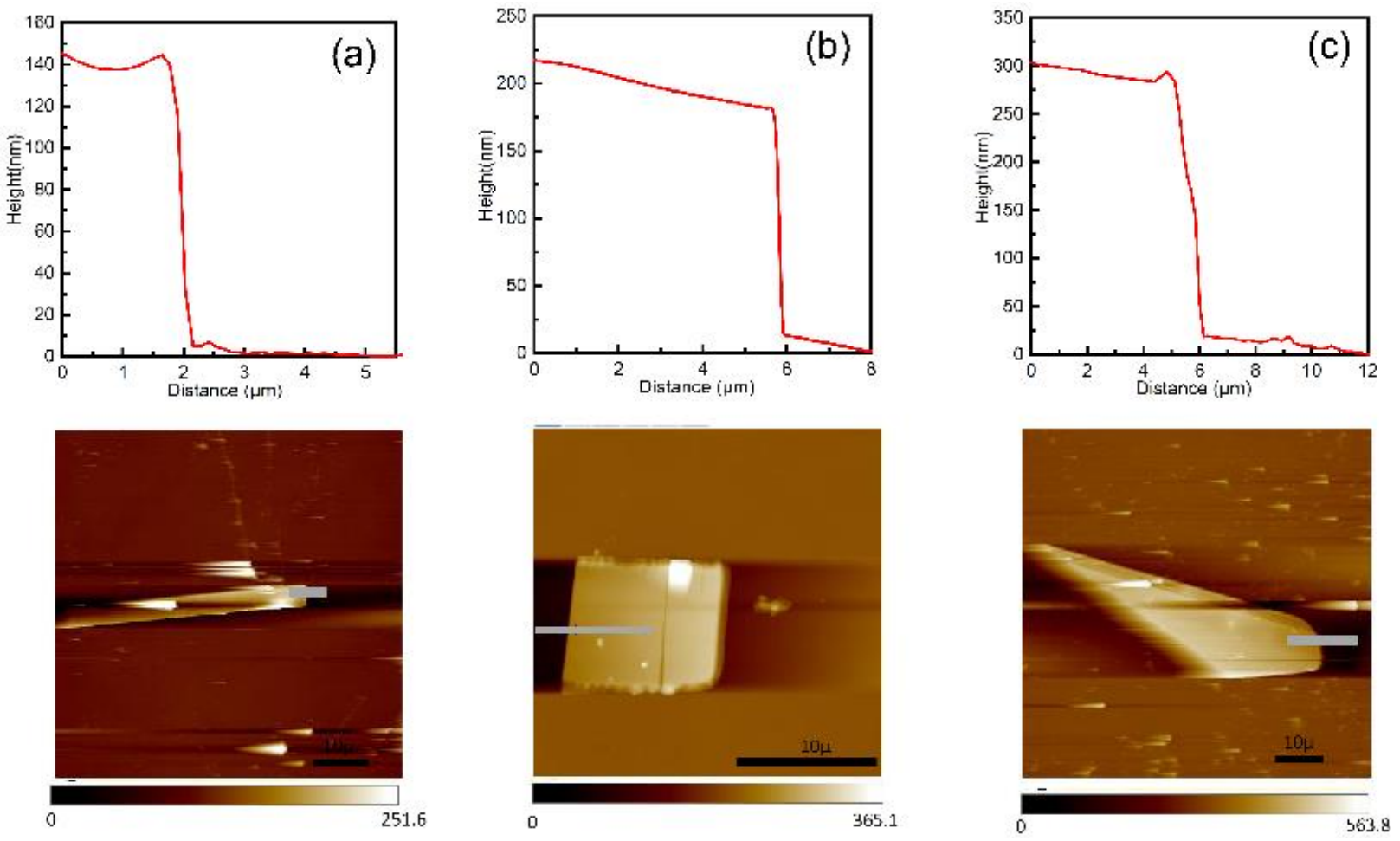

Figure S1: AFM scans on three $MoO_3$ flakes. The lower panel show the real space images of the flake. The upper panel show the corresponding line scans along the grey line depicting the height of the flakes.

## 2. Angles determination in Raman studies

This work presents three different types of Raman studies on $MoO_3$ flakes. These include (i) Raman anisotropy, (ii) Polarization-resolved Raman and (iii) Helicity-resolved Raman. Each of these measurements involve a series of measurement where the angle of polarization in the detection is gradually changed. However, determination of these angles is slightly different in each of these cases. In this part, we fully explain the experimental procedure of angle determination in each case.

### (i) Angle Determination in Raman Polarization Anisotropy

Raman polarization anisotropy studies on MoO3 flakes show in Figure 2 are measured using the setup configuration shown in Fig. The linear polarizer in the incident beam and that in the detection path are initially cross-polarized. A half waveplate (HWP) in the incident path is inserted and rotated until the cross-polarized state of the laser in the detection is restored. This determines the crystal axis of the HWP in the incident beam. A similar procedure is performed for the HWP in the detection path. The $MoO_3$ flake is fixed below the objective such that the incident beam is linearly polarized, where the polarization axis is defined by the polarizer placed in the incident beam. For the data in Figure 2, the sample is kept fixed while the HWP is gradually rotated from its crystal axis. This gradually rotates the linear polarization of the incident laser with respect to crystal axes of the $MoO_3$ flake. The HWP in the detection path is simultaneously rotated in the same direction as that in the incident beam, which ensures detection of the polarization component of scattered Raman signal parallel to the incident laser polarization. Figure 2(d)-(h) shows the normalized intensity of five Raman modes as a function of the incident in-plane polarization orientation with respect to the crystal axis. Zero HWP angle is defined such that the incident polarization is aligned along the crystallographic axis. This is the angle at which the $A_g$ mode at 364 $cm^{-1}$ has minimum intensity.

(ii) Angle determination in polarization-resolved Raman

Figure 3 shows the result of polarization resolved Raman. In this case, the incident polarization is fixed while the detection polarization is gradually rotated in-plane. Using the Raman anisotropy measurements, we identify the orientation of HWP in the incident beam for which the incident laser polarization is along the crystallographic c-axis. Polarization resolved Raman spectra are taken under this particular excitation condition by gradually rotating the HWP in the detection arm.

(iii) Angle determination in helicity resolved Raman measurements

Linear polarized laser is prepared using a polarizer in the incident beam. The detection arm only consists of a cross-polarized linear polarizer. The sample is replaced with a metallic mirror, as the mirror preserve the polarization of the laser. The crystal axis of the QWP is determined by rotating it until the laser the cross-polarization state of laser in the detection arm is restored. The QWP is then rotated by +45 degrees or -45 degrees to set it to either σ+ or σ- polarization. The sign of circular polarization state is determined by calculating the Stokes parameters of laser below the objective. We use the method described in Reference S1 for measuring the Stokes parameter of laser falling on the sample.

Once the QWP angles for incident circular polarization are set, a half wave plate (HWP) is placed within the detection path and a mirror placed at the sample position (see Fig. of the main text). The incident laser is set at a particular polarization state, let's say σ+, using QWP. The HWP is rotated until maximum intensity of laser passes through the last polarizer before the spectrometer. This angle is assigned as the co-polarized angle (or σ+/σ+). A 45-degree rotation from co-polarized configuration of the HWP result in minimum laser passing through the last polarizer. This angle of HWP is assigned as the cross-polarized angle (or σ+/σ-). The two spectra shown in Figure 4 of the main text are taken at these two configurations. For Figure 5, spectra were taken by gradually changing the orientation of the HWP in the detection path. Zero angle is set for the co-polarized configuration.

## 3. Helicity-resolved Raman on flakes of different thickness

The nature of helicity resolved Raman from MoO3 doesn't change much with the thickness of the flakes. Figure S2 present helicity-resolved spectra on 6 different flakes with thicknesses ranging from 150 nm to 950 nm. These spectra were obtained under σ+ excitation. The helical nature of the modes as described in the main text stays the same. Furthermore, Fig. S3 shows helicity-resolved spectra on the same flake under σ+ excitation.

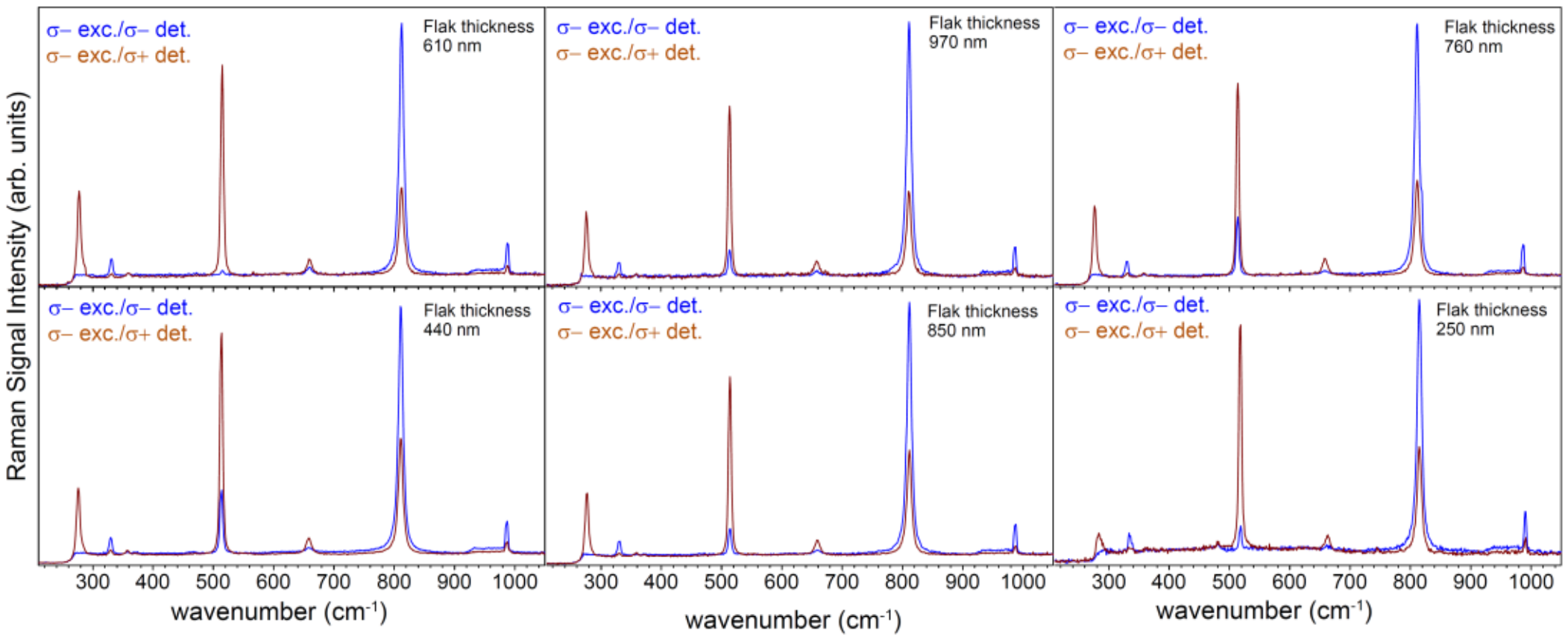

Figure S2: Helicity resolved Raman under σ+ excitation on 6 $MoO_3$ flakes of difference thicknesses. The blue line represents the co-polarized backscattered Raman component (with σ+ excitation and σ+ detection), while the brown line represents the cross-polarized component (with σ+ excitation and σ- detection). Thickness of each flake is mentioned.

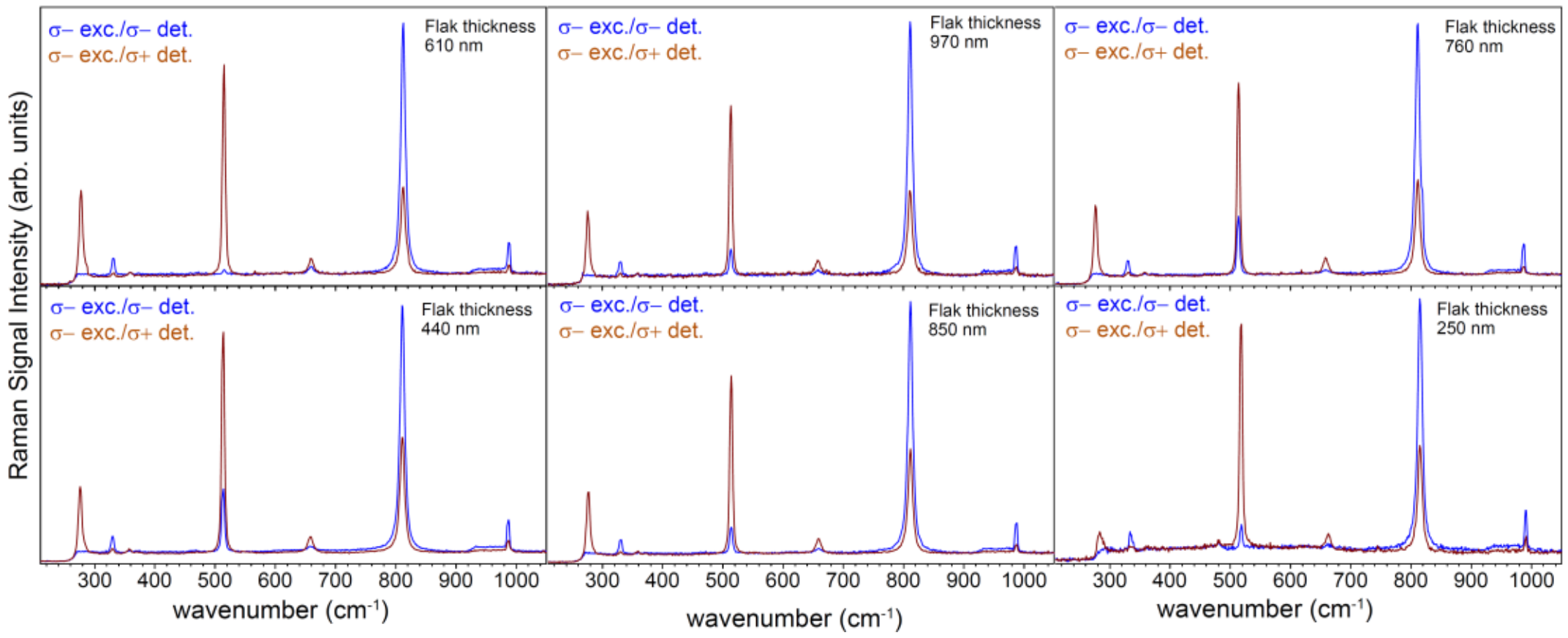

Figure S3: Helicity resolved Raman under σ- excitation on 6 $MoO_3$ flakes of difference thicknesses. The blue line represents the co-polarized backscattered Raman component (with σ- excitation and σ- detection), while the brown line represents the cross-polarized component (with σ- excitation and σ+ detection). Thickness of each flake is mentioned.

## 4. Raman helicity selectivity and sample orientation

Recent experimental studies on α-$MoO_3$ flakes have reported flakes to be in the crystallographic *a-c* plane [S2] or *b-c* plane [S3]. In this study, we used selection rules for Raman scattering to determine the flakes orientation. As mentioned in the main text, Raman scattering intensity depends on $I \propto |e_s \cdot R \cdot e_i|^2$ . The values of $|e_s \cdot R \cdot e_i|^2$ for different combinations of the helical incident and scattered laser is shown in the Table S1. We present data for the cases when the plane of the flake is defined by the crystallographic *a-c* axes and *b-c* axes. Apart from the intensity of $A_g$, the modes $B_{2g}$ and $B_{3g}$ have clear selection rules for both possible orientations of the flakes. Raman scattering of $B_{2g}$ mode is absent in in-plane *b-c* flake orientation, while the $B_{3g}$ mode is absent in *a-c* mode. Comparing these with our experimental data shown in Figure 4 of the main text. The $B_{2g}$ mode at 283 cm$^{-1}$ has higher intensity as compared with the $B_{3g}$ mode at 666 cm$^{-1}$. We infer that the flake is oriented in *a-c* crystallographic plane. The $B_{3g}$ is not completely absent because the incident laser is in the form of a focused beam, so a non-negligible part of the incident beam is incident at non-normal angle on the flake surface. Both the $B_{2g}$ and $B_{3g}$ modes have cross-polarized character as predicted from the data in Table S1.

| | Flake with in-plane *a-c* crystal axes | | | | Flake with in-plane *b-c* crystal axes | | | |
|---|---|---|---|---|---|---|---|---|
| Excitation | $\sigma +$ | $\sigma +$ | $\sigma -$ | $\sigma -$ | $\sigma +$ | $\sigma +$ | $\sigma -$ | $\sigma -$ |
| Detection | $\sigma +$ | $\sigma -$ | $\sigma -$ | $\sigma +$ | $\sigma +$ | $\sigma -$ | $\sigma -$ | $\sigma +$ |
| $A_g$ | $\|a+c\|^2$ | $\|a-c\|^2$ | $\|a+c\|^2$ | $\|a-c\|^2$ | $\|b+c\|^2$ | $\|b-c\|^2$ | $\|b+c\|^2$ | $\|b-c\|^2$ |
| $B_{1g}$ | 0 | 0 | 0 | 0 | 0 | 0 | 0 | 0 |
| $B_{2g}$ | 0 | $4\|e\|^2$ | 0 | $4\|e\|^2$ | 0 | 0 | 0 | 0 |
| $B_{3g}$ | 0 | 0 | 0 | 0 | 0 | $4\|f\|^2$ | 0 | $4\|f\|^2$ |

Table S1: Table showing the $|e_s \cdot R \cdot e_i|^2$ for helical incident beam and helicity resolved detection. These values are shown for two configurations: left side of the table represent calculations for the case when the surface of the flake is defined by the crystallographic a-c plane and the right side of the table represents the case when the surface of the flake is in the crystallographic *b-c* plane.

## 5. Fitting Parameters from Polar Plots

| Mode Center ($cm^{-1}$) | Mode Designation | Fitting Parameters: Raman Anisotropy (Figure 2(d)-(h)) | | | Extracted Parameters: For Helicity Resolved | | |
|---|---|---|---|---|---|---|---|
| | | $a$ | $c$ | $e$ | $\|a+c\|^2$ | $\|a-c\|^2$ | $\|e\|^2$ |
| 283 | $B_{2g}$ | - | - | 0.87 | - | - | 0.76 |
| 336 | $A_g$ | 0.89 | 0.83 | - | 2.96 | 0.003 | - |
| 364 | $A_g$ | 0.96 | -0.04 | - | 0.85 | 1.00 | - |
| 817 | $A_g$ | 0.15 | 0.87 | - | 1.04 | 0.52 | - |
| 992 | $A_g$ | 0.35 | 0.92 | - | 1.49 | 0.27 | - |

Table S2: Figure 2(d)-(h) shows polar plots of normalized Raman modes intensity plotted as a function of the angle between incident polarization and crystallographic *c*-axis. The data points are fitted using Equations (4) and (5) of the main text. This table shows the parameters of Raman tensor a, c and e, as extracted from the fitting for the 5 Raman modes. In the right column, we use these parameters to predict the relative intensities of co- and cross-polarized components as shown in Eqs. (7)-(10) of the main text.

## Supplementary Material References

[S1] Berry, H. Gordon, G. Gabrielse, and A. E. Livingston. "Measurement of the Stokes parameters of light." Applied optics 16 (12), 3200-3205 (1977).

[S2] Wen, M., Chen, X., Zheng, Z., *et al*. "In-Plane Anisotropic Raman Spectroscopy of van der Waals α-$MoO_3$." The Journal of Physical Chemistry *C* **125**, 765-773 (2020).

[S3] Zheng, B., Wang, Z., Chen, Y., *et al*. "Centimeter-sized 2D α-$MoO_3$ single crystal: growth, Raman anisotropy, and optoelectronic properties." 2D Materials **5**, 045011 (2018).